# Threshold Switching in CdTe Photovoltaics

Suman Devkota[a], Kwame A Nyako[a], Brendan Kuzior[a], Victor G. Karpov[b], Daniel G. Georgiev[c], Frank X. Li[a], Pedro Cortes[d], Vamsi Borra[a*]

[a]Electrical and Computer Engineering, Youngstown State University, Ohio
[b]Department of Physics and Astronomy, University of Toledo, Ohio
[c]Department of Electrical Engineering and Computer Science, University of Toledo, Ohio
[d]Chemical Engineering, Youngstown State University, Ohio
*Corresponding Author: vsborra@ysu.edu

With the ubiquitous acceptance of PVs, the number of devices manufactured annually is following an exponential trend. Yet, the manufacturing process of important brands of thin-film solar cells involves a tedious and expensive step of laser scribing. The time-consuming and technologically involved laser scribing method remains widely used to contact the device electrodes. This work examines an alternative method (threshold switching phenomenon) to create an enduring conductive path in cadmium telluride (CdTe) PV, which eliminates the pitfalls in conventional scribing technology. The samples undergone threshold switching show a promising sign of the conductive path compared to the control samples. This method could potentially lead to the manufacturing technology saving time, money, and raw materials along with the added reliability and efficiency.

## Introduction

Thin film photovoltaics (PVs) is a major solar PV brand that offers low consumption of materials and higher mechanical flexibility[1]. Among all other thin films, Cadmium Telluride (CdTe) based PV became most popular with installations possessing close to ideal band gap of 1.45 eV[2]. Important components of CdTe PV research and development include robust film growth, understanding the device physics, and stability issues [3]. Fewer studies have been conducted on PV module integration, which is important to utilize its constituting cells efficiency [4] [5].

Manufacturing of a typical thin film PV module includes a step of laser scribing to establish a contact with the buried electrode. That step makes the production of PVs more expensive, and requiring additional resources[6], [7]. As illustrated in Figure 1, laser scribing along with sequential material deposition is employed to establish desired pathways for electric currents. Various laser scribing schedules are used during the module fabrication including the P3 process [8]of 'type 1' or 'type 2'. The laser scribing performed to remove the layers exposing the back contact is P3 'type 1', whereas the removal of front contact constitutes the P3 of 'type 2'[8]. Figure 1 represents the schematic view of P1, P2 and P3 laser scribing processes.

Recently it was found that the thin film solar cell exhibits the phenomenon of threshold switching qualitatively similar to that of phase change [9] and resistive memory; the switching effect was efficient under forward bias [10][11].Threshold switching is a well-controlled process causing a drastic decrease in resistance under electric field exceeding a certain threshold value[10]. Our study here is focused on experimentally verifying the threshold switching phenomenon in CdTe PV and record possible changes in the switched resistance value as a function of time. The practical significance of our study is that it can pave a way to the scribe less technology in PV.

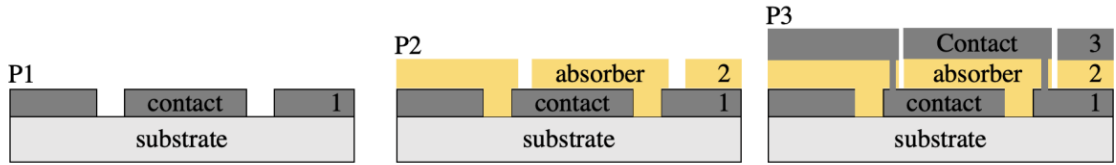

Figure 1. Industrial laser scribing process P1, P2 and P3. Isolation created to a contact using scribing process (P1). Scribe done after deposition of absorber (P2). Scribe done to expose the back contact by removing front contact and absorber (P3). Interconnect between contact 3 and contact 1, in P3, performed using laser scribing enables the extraction of the charges that are collected at the back contact.

## Method

*Threshold Switching Experiment*

The experiment was conducted on a CdTe PV samples shown in Figure 2(b). These sample were prepared by two different techniques: vapor transport deposition and radio frequency magnetron sputtering. In both cases, a layer of CdS followed by CdTe was deposited on commercially available $SnO_2$-coated glass substrates. The transparent conductive oxide layer served as a front electrode. After deposition, the samples were submitted to a standard anneal in the presence of $CdCl_2$ vapor which generally leads to improved electrical characteristics. Finally, a metal layer was deposited to form the back contact to CdTe [12]. The prepared sample was properly handled to prevent it from damaging physically and chemically by following standard safety protocols. Figure 2 shows the schematic of the CdTe PV, picture of the actual CdTe sample used, and 50X magnified image of an arbitrary scribe. The experiment was performed under room temperature on a 10 x 10 cm CdTe PV sample figure 2(b). As shown in the figure 2(c) small scribe of approximately 3.5mm was created. A probe station, with micromanipulator and pogo pins, was used to measure the resistance and supply the required voltages without damaging the sample. Programable DC power supply and digital multimeter were employed to supply voltage and measure resistance respectively. The voltage was applied across a point inside the scribe and a point in TCO, and resistance was measured between those two points.

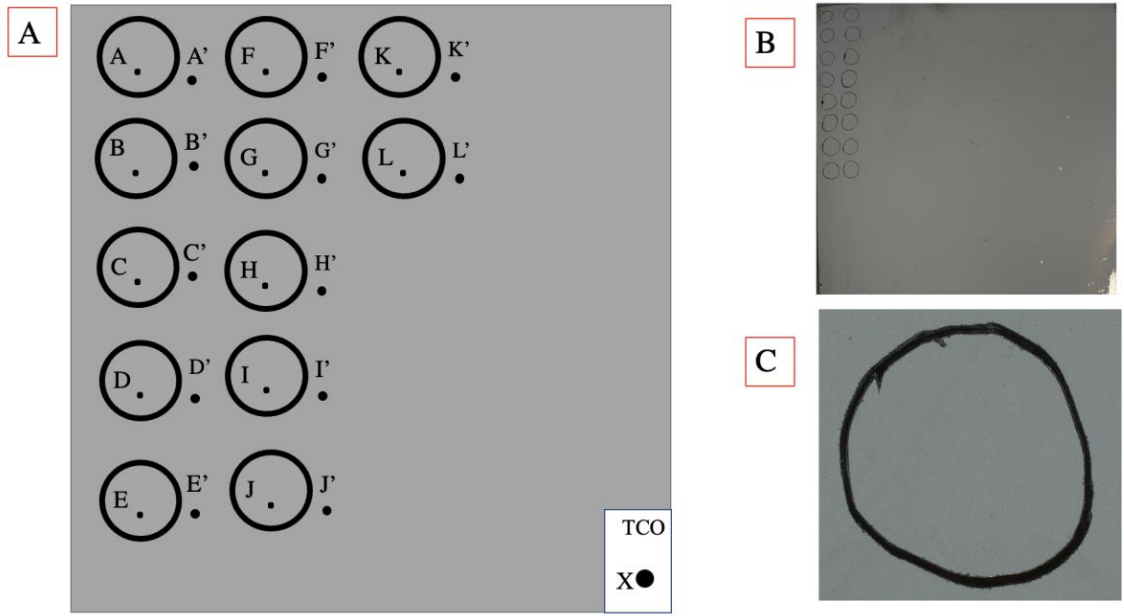

Figure 2. Schematic of the CdTe PV, points(A-L) are the points inside different scribes; point X is the point on transparent conductive oxide (TCO) (A). Sample used for the experiments (B), and x50 magnified image of a sample with a scribe of 3.5mm (C).

In summary, the following steps are followed to conduct the threshold switching experiment:
I. First, a reference resistance between an arbitrary point, outside the scribe, and TCO is recorded i.e., Resistance(A'-X) for scribe A, Resistance(B'-X) for scribe B and others (see figure 2 (a)). These values were recorded to compare the resistance value before and after the voltage bias.
II. Scribes *A, B, C, D, E and F* were supplied with 0V, 5V, 10V, 15V, 20V and 25V respectively for 10 seconds.
III. Similarly, another scribe set, *G, H, I, J, K* and *L* were supplied with 0V, -5V, -10V, -15V, -20V, and -25V respectively for 10 seconds.

The above steps were repeated multiple times on separate samples to verify the results.

*Time-lapse study of switched resistance*

The other part of the experiment is to understand the reliability of threshold switching and endurance of its trait over the time. After the threshold switching experiment, the resistance of the created shunt, across the point inside the scribe and on a point in TCO, was recorded after 6, 12, 24, and 48 hours.

## Results and Discussion

Threshold switching

The studied samples show drastic decrease in resistance, across the point inside the scribe and on a point in TCO, ranging from MΩ to kΩ when voltage biased. A similar effect, a strong decrease in resistance, was observed when the negative voltage bias was supplied as well. This sharp decrease in the resistance, post voltage bias, implies that there is a creation of a conductive shunt through the point on the scribe reaching the buried electrode TCO for this sample. Figure 4 shows the decreased value of resistance when different voltages were supplied. **Table I** and **Table II** show the resistance value recorded before and after supplying positive and negative voltages respectively.

TABLE I. Change in resistance after supplying positive voltage for 10 seconds.

| Scribe | Resistance across scribe and TCO prior bias | Supply Voltage(V) | Resistance across scribe and TCO after bias |
|---|---|---|---|
| A | 1.07E+06 Ω | 0 | 1.06E+06 Ω |
| B | 1.02E+06 Ω | 5 | 1.13E+05 Ω |
| C | 1.05E+06 Ω | 10 | 4.92E+04 Ω |
| D | 1.00E+06 Ω | 15 | 3.49E+03 Ω |
| E | 1.01E+06 Ω | 20 | 1.42E+04 Ω |
| F | 1.02E+06 Ω | 25 | 1.62E+04 Ω |

TABLE II. Change in resistance after supplying negative voltage for 10 seconds.

| Scribe | Resistance across scribe and TCO prior bias | Supply Voltage(V) | Resistance across scribe and TCO after bias |
|---|---|---|---|
| F | 1.02E+06 Ω | 0 | 1.02E+06 Ω |
| G | 1.77E+06 Ω | -5 | 1.10E+05 Ω |
| H | 1.08E+06 Ω | -10 | 1.10E+05 Ω |
| I | 1.07E+06 Ω | -15 | 1.09E+05 Ω |
| J | 1.01E+06 Ω | -20 | 1.63E+03 Ω |
| K | 1.02E+06 Ω | -25 | 3.00E+03 Ω |

The sequence of schematics in Figure 3 show (a) the cross-sectional view of the CdTe PV showing the sandwich structure, (b) sample after scribing with a sharp object show an island like structure created, (c) red arrows signify the voltage bias to the island and the TCO from the power supply to form the shunt, (d) red solid conductive shunt was formed after leaving the power supply for 10 seconds across the island and TCO, signifying threshold switching.

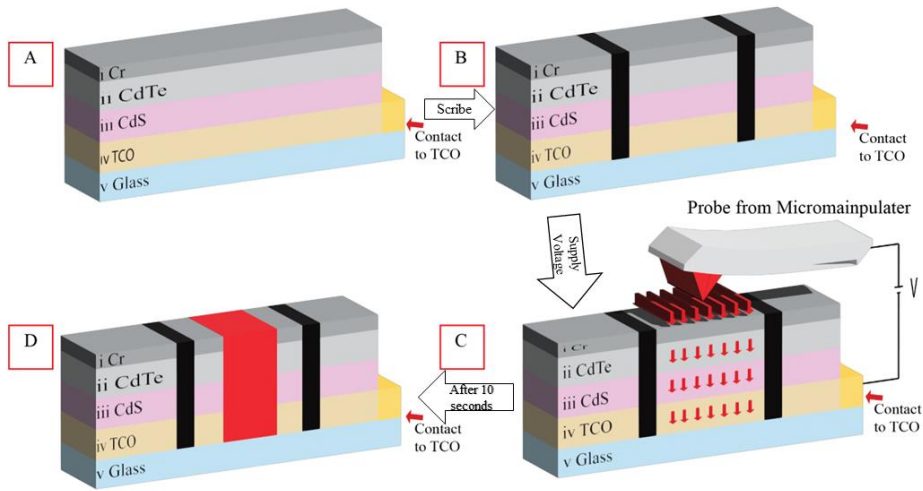

Figure 3. Cross-sectional view of the CdTe (not to scale) photovoltaics. Starting from top i) thin chromium layer ii) CdTe iii) CdS iv) glass layer (A). Cross-section of same sample after scribing (B). Red arrows indicating supply voltage to the scribe (C). Red rectangle indicating creation of a conductive shunt (D).

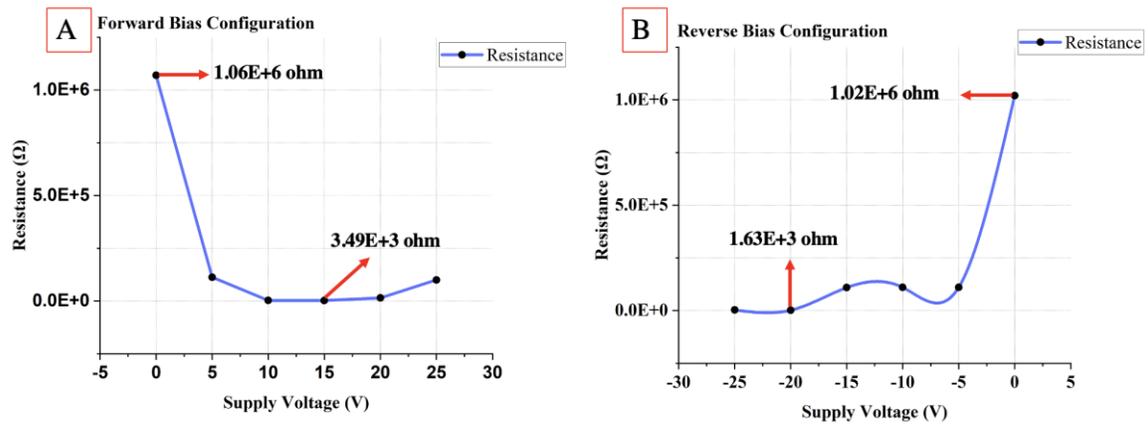

Figure 4. Rapid drop in resistance value with positive voltage bias configuration, scribes A to F (A) and negative voltage bias configuration, scribes G to L (B).

Resistance across the individual scribes and TCO after the bias value, tabulated in Table I and II, are plotted in the Figure 4. It can be noted that the resistance change after the positive voltage bias is quite significant across the scribes A to F, Figure 4 (A). As detailed in the above Table I, a voltage increments of 5 V, starting from 0 V to 25 V, biased at each scribe for 10 seconds verified the formation of a conductive path. Similarly, similar phenomenon is observed when a negative voltage bias is applied across the scribes (G to L) and TCO, Figure 4 (B). This observed change, from MΩ to kΩ, could potentially enable the scribe less technology by eliminating the tedious laser scribing.

Reliability of threshold switching

The time-lapse study, performed to check the reliability of the formed conductive shunt, indicated that the formed shunt's resistance value was in the ballpark of 2 KΩ. Figure 5 (A) shows the change in resistance observed over time for positive voltage bias, Scribe D whereas figure 5 (B) shows the change on resistance for negative voltage bias, Scribe I. **Table III** shows the resistance value recorded over the period of 6, 12, 24, 48 hours. Here, 0 hours indicate the time when the initial shunt was formed by the voltage bias.

**TABLE III.** Shunt's resistance value over period

| Scribe | 0 hour | 6 hours | 12 hours | 24 hours | 48 hours |
|---|---|---|---|---|---|
| C | 2.87E+03 Ω | 2.80E+03 Ω | 4.08E+03 Ω | 3.39E+03 Ω | 3.07E+03 Ω |
| D | 2.51E+03 Ω | 2.06E+03 Ω | 4.20E+03 Ω | 3.39E+03 Ω | 3.11E+03 Ω |
| I | 4.64E+04 Ω | 4.80E+04 Ω | 5.11E+04 Ω | 5.10E+03 Ω | 5.07E+04 Ω |
| K | 1.68E+03 Ω | 1.96E+03 Ω | 2.80E+03 Ω | 2.94E+03 Ω | 2.07E+03 Ω |

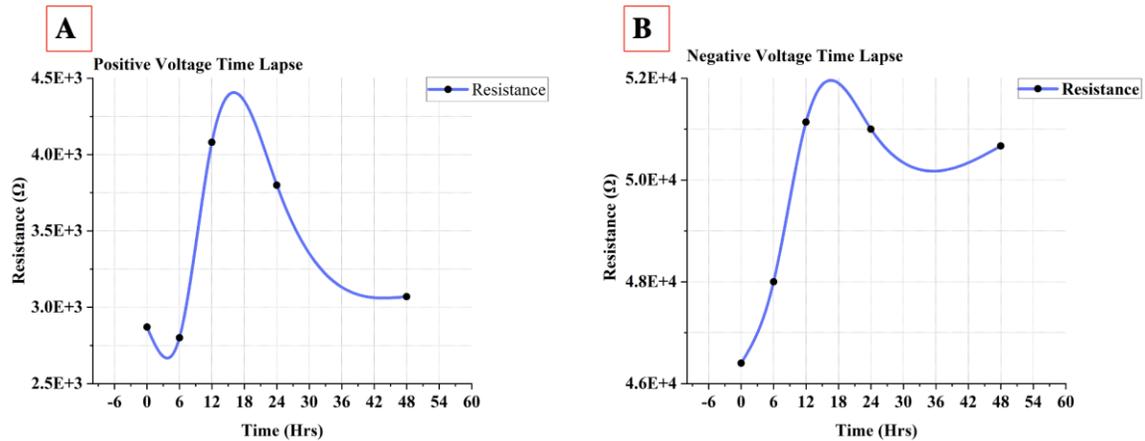

Figure 5. CdTe PV change in resistance is consistent over a period of time. Change in resistance where supply voltage was 10V, (Scribe D) (A) and change in resistance where supply voltage was -10V, (Scribe I) (B).

The observed switching mechanism could have been resulted due to the crystallization or rearrangement of specific structural imperfections, impurities, or electric dipoles into a narrow conducting filaments, which were previously detected [13]–[15]. In addition, a similar phenomenon was observed by redox (reduction/oxidation) effect [16]. According to that theory, the switching is a result of electrochemical reactions involving the mobility of ions and oxygen vacancies at the ionic level, which could be a reason for the formation of the conductive filament. Finally, nucleation theory [17] that associate nucleation rate and applied voltage could explain the above observation.

## Conclusion

In this study, an enhanced method to create an enduring conductive path in cadmium telluride (CdTe) PV was proposed. This method could be an alternative to the existing time-consuming and technologically involved laser scribing method. Laser scribing, a process step in the thin film PV module manufacturing, enables the contact with the buried electrode. This step requires additional resources and can lead to increase in the fabrication costs. Our experimental results imply that the voltage bias creates a conductive filament from the top most layer through the PV stack to the buried electrode, through the phenomenon of threshold switching. Time lapse study of the conductive filament's resistance suggest that the threshold switching phenomenon is reliable. This novel method could potentially lead to scribe less technology of manufacturing PV by elimination the back contact.

With the study and experiment conducted above we are able to conclude the following:
- Thin film of CdTe PV exhibit the phenomenon of threshold switching.
- Switching was observed on both positive and negative voltage configuration.
- Switching effect lasted longer implying its reliability.
- Threshold switching phenomenon could potentially lead to scribe less technology by eliminating the conventional tedious and expensive of laser scribing.

## Acknowledgments

The authors would like to thank the U.S. Air Force Research Laboratories under The Assured Digital Microelectronics Education and Training Ecosystem (ADMETE) project for providing financial support.